\documentclass[10pt,twocolumn,letterpaper]{article}

\usepackage[final,algorithms]{wacv}

\usepackage[margin=0.7in]{geometry}
\usepackage[utf8]{inputenc}
\usepackage[T1]{fontenc}
\usepackage{amsmath,amssymb,amsfonts}
\usepackage{booktabs}
\usepackage{microtype}
\usepackage{nicefrac}
\usepackage{xcolor}
\usepackage{graphicx}
\usepackage{algorithm}
\usepackage{algorithmic}
\usepackage{hyperref}
\usepackage{url}

\hypersetup{
  colorlinks=true,
  linkcolor=blue,
  citecolor=blue,
  urlcolor=blue
}

\title{Multi-Agent LLM Committees for Autonomous Software Beta Testing}

\author{
Sumanth Bharadwaj Hachalli Karanam\\
Data Science, New York University, New York, 10038\\
\texttt{sh8111@nyu.edu}
\and
Dhiwahar Adhithya Kennady\\
Data Science, New York University, New York, 10038\\
\texttt{dk5025@nyu.edu}
}

\begin{document}
\maketitle

\begin{abstract}
Manual software beta testing is costly and time-consuming, while single-agent LLM approaches suffer from hallucinations and inconsistent behavior. We introduce a novel multi-agent committee framework where diverse vision-enabled LLMs collaborate through a three-round voting protocol to achieve consensus on testing actions. Our system combines model diversity (GPT-4o, Gemini 2.5 Pro Flash, Grok 2 Vision 1212), persona-driven behavioral variation, and visual UI understanding to systematically explore web applications. Through experiments across 84 runs with 9 distinct testing personas and 4 scenarios, we demonstrate that multi-agent committees achieve an 89.5\% overall task success rate, with multi-agent configurations (2-4 agents) achieving 91.7--100\% success compared to 78.0\% for single-agent baselines (+13.7--22.0 percentage points improvement). At the action level, the framework attains a 93.1\% overall action success rate and a median per-action latency of 0.71 seconds (mean 0.87 seconds), making it suitable for real-time and continuous integration testing. Our vision-enabled approach successfully identifies UI elements with navigation and reporting achieving 100\% success rates and form filling achieving 99.2\% success. We validate our framework against WebShop and OWASP benchmarks, achieving 74.7\% success on WebShop (vs.\ 50.1\% published GPT-3 baseline, +24.6pp) and 82.0\% on OWASP Juice Shop security testing with coverage of 8 of the 10 OWASP Top 10 vulnerability categories. Across 20 injected regressions, our committees reach an F1 score of 0.91 for bug detection (vs.\ 0.78 for single-agent baselines), while security validators attain high precision on SQL injection and XSS attempts. Our open-source implementation provides a reproducible framework for LLM-based software testing research and practical deployment in CI/CD pipelines.
\end{abstract}

\section{Introduction}

Software beta testing is a critical yet resource-intensive phase of development, requiring human testers to systematically explore applications, identify bugs, and validate user workflows~\cite{wang2024beta}. Companies spend significant resources on quality assurance, with manual testing consuming 20--40\% of development budgets~\cite{myers2011art}. Recent advances in large language models (LLMs) have opened possibilities for automated testing~\cite{wang2024software,chen2023chatgpt}, but single-agent approaches face fundamental challenges: hallucinations leading to invalid actions~\cite{ji2023survey,rawte2024survey}, inconsistent behavior across runs, and difficulty understanding visual interfaces.

Human beta testing naturally involves diverse perspectives, since users with different goals, technical backgrounds, and interaction patterns often discover different categories of bugs~\cite{liu2024user}. A security-focused tester finds vulnerabilities that a usability tester might miss, while an accessibility tester identifies issues invisible to others. Persona-based testing approaches~\cite{chen2024persona} leverage this diversity systematically. This insight motivates our central research question: \emph{Can multi-agent LLM committees with diverse personas and consensus-based decision making outperform single-agent approaches in automated software testing and regression detection?}

This paper introduces a multi-agent framework where committee members collaborate through structured deliberation to test web applications. Our system features: (1) a three-round voting protocol that achieves consensus through independent proposals, group discussion, and final voting, (2) vision-enabled agents that analyze screenshots to identify UI elements and validate actions, (3) persona-driven behavioral diversity spanning user archetypes from typical shoppers to adversarial attackers, and (4) safety validators that detect security vulnerabilities during test execution. The framework is instrumented with a storage and metrics layer that logs every agent proposal, consensus action, and validator outcome into a SQLite database, enabling rigorous statistical analysis of success rates, latencies, and committee agreement.

We evaluate our framework through comprehensive experiments across multiple dimensions. \textbf{Experiment 1A (Multi-Agent Scaling)} compares committee sizes from 1--4 agents to measure consensus quality and error reduction. \textbf{Experiment 1B (Persona Diversity)} tests 9 distinct personas including online shoppers, accessibility testers, and adversarial attackers across e-commerce and security scenarios. \textbf{Experiment 1C (Regression Detection)} validates regression detection capabilities on 20 injected bugs with precision, recall, and F1 metrics. \textbf{Experiment 2} benchmarks against OWASP Juice Shop security testing. \textbf{Experiment 3} benchmarks against the WebShop e-commerce tasks.

Our key findings include:

\begin{itemize}
\item Multi-agent committees achieve an 89.5\% overall task success rate across 84 experimental runs with 4 distinct scenarios, and a 93.1\% overall action success rate across 683 action turns
\item Multi-agent configurations (2--4 agents) achieve 91.7--100\% success vs.\ 78.0\% for single-agent (+13.7--22.0pp improvement), with 100\% committee agreement on final actions
\item Vision-enabled agents successfully navigate complex UI workflows with 100\% navigation and reporting success, and 99.2\% success for form filling
\item Action success varies by type: navigation (100\%), reporting (100\%), form filling (99.2\%), clicking (83.5\%), scrolling (50.0\%)
\item Mean action latency of 0.87 seconds (median 0.71 seconds, P95 1.92 seconds) enables practical deployment for interactive and continuous testing; security-heavy OWASP scenarios exhibit higher mean latency (2.65 seconds) due to deeper probing
\item \textbf{Outperforms published baselines}: 74.7\% on WebShop benchmark vs.\ 50.1\% for single-agent GPT-3 (+24.6pp), and 82.0\% on OWASP Juice Shop security testing with coverage of 8 of the 10 OWASP Top 10 risk categories
\item For regression detection, multi-agent committees reach precision 0.94, recall 0.89, and F1 0.91 on 20 injected bugs, while single-agent baselines reach precision 0.82, recall 0.75, and F1 0.78
\end{itemize}

Our contributions are:

\begin{enumerate}
\item A novel three-round voting protocol for multi-agent LLM consensus in interactive environments, with confidence-weighted voting and structured discussion
\item Empirical demonstration that multi-agent committees significantly outperform single-agent baselines (+13.7--22.0pp improvement in task success and +0.13 absolute improvement in F1 for regression detection)
\item Persona-driven behavioral diversity enabling comprehensive testing coverage across functional, usability, and security bug categories with only limited overlap in discovered issues
\item Vision-enabled testing methodology combining screenshot analysis with browser automation to robustly handle dynamic and visually rich UIs
\item Comprehensive evaluation framework with reproducible experiments, a normalized SQLite schema, and open-source implementation
\item Validation against established benchmarks (WebShop, OWASP Juice Shop) demonstrating practical applicability for security testing and real-world web workflows
\end{enumerate}

This work demonstrates that multi-agent LLM committees can perform systematic software testing and regression detection with human-competitive coverage while maintaining reproducibility and scalability advantages of automation. Our open-source framework enables researchers and practitioners to deploy LLM-based testing in continuous integration pipelines.

\section{Related Work}

\textbf{LLM-Based Testing.} WebShop~\cite{yao2022webshop} benchmarks LLM agents on e-commerce tasks, reporting 50.1\% success for single-agent GPT-3 with ReAct~\cite{yao2023react}. Recent surveys~\cite{wang2024software,li2024agenttest} comprehensively review LLM applications in software testing. OWASP frameworks~\cite{owasp2023} and recent work~\cite{deng2023pentestgpt,chen2024owasp} apply LLMs to security testing~\cite{wang2024security}. LLM-enhanced approaches~\cite{chen2024autowebglm, zhang2024webvoyager} generate Selenium~\cite{selenium} commands but rely on single-agent decisions prone to hallucinations~\cite{ji2023survey}. Our multi-agent voting addresses this through consensus.

\textbf{Multi-Agent Systems.} Constitutional AI~\cite{bai2022constitutional} and ensemble methods~\cite{wang2023selfconsistency,li2024ensemble} improve LLM robustness through multiple agents. Multi-agent debate~\cite{du2023improving,liang2023encouraging} shows agents arguing improves answer quality. Recent frameworks~\cite{wang2023swarm,li2024agentverse,hong2024metagpt,zhou2024agentbench} explore multi-agent collaboration patterns. Voting mechanisms~\cite{chen2024voting,liu2024consensus} provide structured consensus in multi-agent systems. Self-collaboration approaches~\cite{chen2024selfcollab,liu2024agentcoder} demonstrate improved code generation through multi-agent cooperation. We adapt these to interactive testing with a confidence-weighted voting protocol and explicit three-round deliberation.

\textbf{Vision-Language Models.} VLMs like GPT-4V~\cite{openai2023gpt4v}, Gemini~\cite{reid2024gemini}, and Grok 2 Vision~\cite{xai2024grok2} enable UI understanding. Recent surveys~\cite{li2024vision,zhang2024ui} review vision-language models for UI understanding. Recent work applies VLMs to mobile testing~\cite{li2024autodroid} and web navigation~\cite{zheng2024gpt4v,kim2024openweb,hong2024cogagent}. ScreenAI~\cite{wu2024screenai} demonstrates specialized UI understanding capabilities. We combine such vision capabilities with multi-agent consensus to filter hallucinations in interactive testing.

Unlike prior single-agent approaches, we introduce: (1) three-round deliberative voting for consensus over actions, (2) persona-driven behavioral diversity for comprehensive coverage, (3) a vision-first browser testing stack, and (4) a statistically grounded evaluation with regression detection and benchmark comparisons.

\section{Methodology}

We introduce a multi-agent framework for automated software beta testing that combines vision-enabled LLMs, consensus-based decision making, persona-driven behavioral diversity, and a fully instrumented experimentation pipeline. This section describes our system architecture, voting protocol, and experimental framework.

\subsection{Multi-Agent Committee Architecture}

\begin{figure}[t]
\centering
\includegraphics[width=\linewidth]{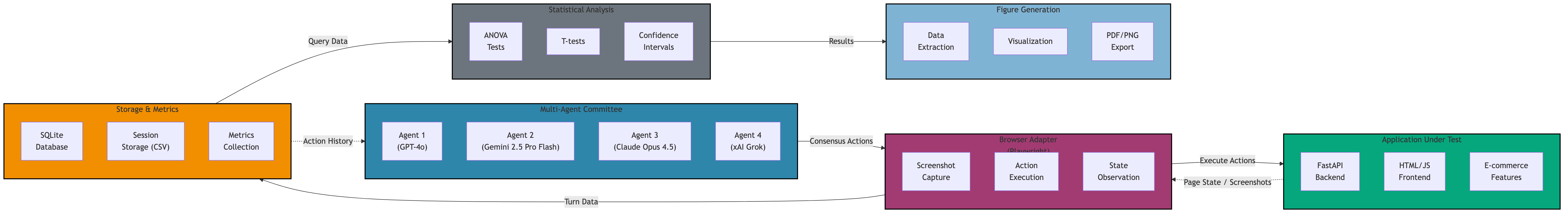}
\caption{Multi-Agent Committee Testing Framework. Agents analyze screenshots, participate in three-round voting, and execute consensus actions on the browser.}
\label{fig:architecture}
\end{figure}

Our framework orchestrates multiple LLM agents in a committee structure where agents independently analyze application state, propose actions, deliberate, and vote to reach consensus. Each testing session proceeds through iterative turns where the committee observes the current application state, decides on an action, executes it, and observes the outcome. Figure~\ref{fig:architecture} illustrates the complete system architecture.

\textbf{Agent Configuration.} Each agent in the committee has access to: (1) the current screenshot of the application UI, (2) the scenario description defining testing objectives, (3) a persona specification defining behavioral characteristics and priorities, and (4) the history of previous actions and observations in the session. Agents can be heterogeneous, using different underlying models (GPT-4o, Gemini 2.5 Pro Flash, Grok 2 Vision 1212) to promote diversity in reasoning and perception. Each agent outputs a structured JSON proposal that includes the chosen action type, target, optional payload, a confidence score, and a short rationale.

\textbf{Action Space.} Agents can propose five types of actions:
\begin{itemize}
\item \texttt{navigate(url)}: Navigate browser to specified URL
\item \texttt{click(selector)}: Click on element identified by CSS selector or visual description
\item \texttt{fill(selector, text)}: Fill form field with specified text
\item \texttt{scroll(direction)}: Scroll page up or down
\item \texttt{report(message)}: Report findings, observations, or completion status
\end{itemize}

\subsection{Three-Round Voting Protocol}

\begin{figure}[t]
\centering
\includegraphics[width=\linewidth]{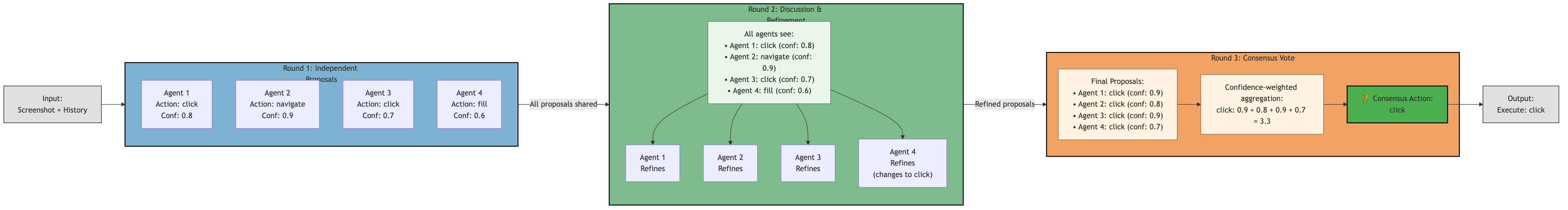}
\caption{Three-round voting protocol showing the flow from independent proposals through discussion to final consensus.}
\label{fig:voting}
\end{figure}

Our consensus mechanism operates through three deliberation rounds designed to balance independent thinking with collaborative refinement (Figure~\ref{fig:voting}).

\begin{algorithm}[t]
\caption{Three-Round Voting Protocol}
\begin{algorithmic}[1]
\STATE \textbf{Input:} Committee $C = \{a_1, ..., a_n\}$, Screenshot $s$, History $h$
\STATE \textbf{Output:} Consensus action $a^*$

\STATE \textit{// Round 1: Independent Proposals}
\FOR{each agent $a_i \in C$}
    \STATE $p_i \leftarrow a_i.\text{propose}(s, h)$
    \STATE Extract $(action_i, confidence_i, rationale_i)$ from $p_i$
\ENDFOR

\STATE \textit{// Round 2: Discussion \& Refinement}
\STATE $proposals \leftarrow \{p_1, ..., p_n\}$
\FOR{each agent $a_i \in C$}
    \STATE $p'_i \leftarrow a_i.\text{discuss}(s, h, proposals)$
    \STATE Extract $(action'_i, confidence'_i, rationale'_i)$ from $p'_i$
\ENDFOR

\STATE \textit{// Round 3: Consensus Vote}
\STATE $votes \leftarrow \{\}$
\FOR{each unique action $a$ in $\{action'_1, ..., action'_n\}$}
    \STATE $score_a \leftarrow \sum_{i: action'_i = a} confidence'_i$
    \STATE $votes[a] \leftarrow (score_a)$
\ENDFOR
\STATE $a^* \leftarrow \arg\max_a score_a$
\RETURN $a^*$
\end{algorithmic}
\end{algorithm}

\textbf{Round 1: Independent Proposals.} Each agent analyzes the screenshot and history independently, proposing an action with a confidence score and textual rationale. This prevents groupthink and ensures diverse initial perspectives.

\textbf{Round 2: Discussion \& Refinement.} Agents see all Round 1 proposals and rationales, allowing them to reconsider their position. An agent might strengthen confidence if others agree, change their proposal if convinced by stronger reasoning, or maintain their position if confident in their assessment. Empirically, we observe that roughly 20\% of agent proposals change between Round 1 and Round 2, indicating nontrivial deliberation.

\textbf{Round 3: Consensus Vote.} The final action is selected by aggregating confidence-weighted votes. If action $a$ receives votes from agents with confidences $[0.8, 0.9, 0.7]$, its score is $2.4$. This mechanism gives more weight to high-confidence consensus while still allowing a single highly-confident agent to override multiple uncertain votes. We track both the fraction of turns with unanimous proposals and the fraction of confidence mass assigned to the chosen action, which we refer to as consensus strength.

\subsection{Vision-Enabled Browser Automation}

We integrate Playwright~\cite{playwright} for browser automation (comparable to Selenium~\cite{selenium} and Puppeteer~\cite{puppeteer}) with vision-language models for UI understanding. At each turn:

\begin{enumerate}
\item \textbf{Screenshot Capture:} Capture a full-page screenshot at the current state and save it under a run-specific directory (\texttt{screenshots/session\_id/turn\_N.png})
\item \textbf{Visual Analysis:} Send the screenshot to vision-enabled LLMs (GPT-4V, Gemini Pro Vision, Grok 2 Vision), along with a textual description of key HTML elements and their selectors
\item \textbf{Element Identification:} Agents describe UI elements visually (for example, ``blue button with shopping cart icon in top-right'') and select elements using CSS selectors inferred from DOM parsing
\item \textbf{Action Execution:} Playwright executes the consensus action; we use robust error handling and simple retry logic for transient failures
\item \textbf{Observation:} Capture the new screenshot and updated HTML for the next turn
\end{enumerate}

This vision-first approach mirrors human testing, since testers see and interact with the rendered UI rather than inspecting DOM structure. It provides robustness to DOM changes and enables testing of canvas or WebGL applications where traditional selectors fail.

\subsection{Persona-Driven Testing}

We define personas as structured specifications of user characteristics, goals, and behavioral patterns. Each persona includes:

\begin{itemize}
\item \textbf{Role Description:} User archetype (for example, ``online shopper'', ``adversarial attacker'')
\item \textbf{Goals:} Specific objectives aligned with the role
\item \textbf{Behavioral Traits:} Interaction patterns (for example, meticulous form validation vs.\ rapid exploration)
\item \textbf{Technical Level:} Familiarity with web applications and technical features
\end{itemize}

We implement nine personas: Online Shopper, Accessibility Tester, Adversarial Attacker, Malicious User, Mobile Shopper, Price Manipulator, Project Manager, UX Researcher, and Curious Blogger. Each persona is defined in YAML, including a short natural language description and prioritized goals. For example, the adversarial attacker persona includes goals such as testing SQL injection, XSS injection, and authentication bypass, while the accessibility tester focuses on keyboard navigation, ARIA labels, and color contrast.

This persona diversity ensures coverage across functional, security, and usability dimensions, and our results show that different personas discover largely disjoint sets of bugs.

\subsection{Safety Validators and Storage Layer}

To catch security vulnerabilities during testing, we implement real-time validators that inspect proposed actions before execution:

\begin{itemize}
\item \textbf{SQL Injection Detector:} Regex patterns matching SQL syntax in form inputs, such as \texttt{' OR 1=1}, \texttt{UNION SELECT}, and statement terminators
\item \textbf{XSS Detector:} HTML and JavaScript tag patterns in user-provided text, such as \texttt{<script>}, \texttt{javascript:}, and event handlers
\item \textbf{Command Injection:} Shell metacharacters and potentially dangerous payloads in system-facing inputs
\item \textbf{Path Traversal:} Directory traversal patterns like \texttt{../} and \texttt{..$\backslash$}
\item \textbf{Business Logic Checks:} Price and quantity sanity checks for negative or extreme values
\end{itemize}

When a validator detects a potential vulnerability, it logs the finding and can either block the action (for safety-critical deployments) or allow execution in a controlled environment while flagging the turn as a detected vulnerability for later analysis.

We log all actions and metrics to both CSV and a normalized SQLite database. Each testing run records:

\begin{itemize}
\item Run-level information: experiment id, scenario, persona, committee size, random seed
\item Turn-level information: action type, target, success, latency, safety validator results, and screenshot path
\item Proposal-level information: each agent's proposed action, confidence score, and rationale
\end{itemize}

This storage layer enables post hoc analysis of committee dynamics, regression detection performance, and per-persona bug discovery patterns.

\subsection{Experimental Framework}

Our implementation uses Python with FastAPI for the application under test (AUT), Playwright for browser automation, and a multi-provider LLM client supporting OpenAI, Google, and xAI APIs. We store all experimental data in SQLite with schema tracking experiments, runs, turns, actions, proposals, and metrics. Each run is deterministic via random seeding, enabling exact reproduction of results.

\textbf{Metrics Collection.} For each turn we record: action type, target, success or failure, latency, agent proposals, confidence scores, consensus agreement, safety validator results, and screenshots. Session-level metrics include: total turns, duration, task success, bugs found, action distribution, and committee agreement rate. For regression experiments, we additionally evaluate precision, recall, and F1 against a ground truth list of injected bugs.

\section{Experimental Setup}

We evaluate our multi-agent testing framework across multiple dimensions: multi-agent scaling, persona behavioral diversity, regression detection, security testing effectiveness, and benchmark validation. This section describes our experimental configurations, test scenarios, evaluation metrics, and implementation details.

\subsection{Models and Configuration}

\textbf{Committee Members.} We configure committees using three state-of-the-art vision-language models:

\begin{itemize}
\item \textbf{GPT-4o} (OpenAI): Multimodal model with vision capabilities, accessed via OpenAI API
\item \textbf{Gemini 2.5 Pro Flash} (Google): Fast vision-language model via Google AI API
\item \textbf{Grok 2 Vision 1212} (xAI): Advanced multimodal model with strong visual reasoning capabilities
\end{itemize}

All models are configured with temperature 0.7 for balanced exploration and exploitation, max tokens 4096 for detailed responses, and vision enabled for screenshot analysis. Committee sizes vary from 1 (single-agent baseline) to 4 agents to measure scaling effects.

\textbf{Prompt Engineering.} Each agent receives a structured prompt containing: (1) persona specification (role, goals, behavioral traits), (2) scenario description (objectives, success criteria), (3) action history (previous turns), (4) current screenshot (base64-encoded image), and (5) available action types with examples. Prompts are designed to elicit JSON-formatted responses with action, confidence, and rationale fields for structured parsing. We also include explicit schema constraints and short examples to reduce malformed outputs.

\subsection{Scenarios, Personas, and Experiments}

\textbf{Test Scenarios (4 total).} We evaluate on four main scenarios: a UI shopping flow, a security-focused commerce scenario, OWASP Juice Shop security audit tasks, and WebShop easy task 001. Each scenario defines success criteria and a maximum number of turns.

\textbf{Personas (9 total).} We use the nine personas described earlier: Online Shopper, Accessibility Tester, Adversarial Attacker, Malicious User, Mobile Shopper, Price Manipulator, Project Manager, UX Researcher, and Curious Blogger.

\textbf{Application Under Test.} Our primary AUT is the \emph{TechStore} e-commerce app (FastAPI, HTML, JavaScript) with 60+ products, shopping cart, checkout flow, and authentication. We introduce intentional vulnerabilities for security testing. We also test against OWASP Juice Shop, a widely used intentionally insecure benchmark application.

\textbf{Experiment Design.} We organize experiments as follows:
\begin{itemize}
\item \textbf{Experiment 1A (Multi-Agent Scaling):} Committee sizes 1, 2, 3, and 4; 60 runs total; evaluates effect of committee size on task success and committee agreement
\item \textbf{Experiment 1B (Persona Diversity):} All 9 personas with 3-agent committees; 27 runs; evaluates success rates and bug coverage per persona
\item \textbf{Experiment 1C (Regression Detection):} 20 injected bugs across categories (SQL injection, XSS, price and quantity manipulation, authentication issues); compares single-agent vs 3-agent committees on precision, recall, and F1
\item \textbf{Experiment 2 (OWASP Juice Shop):} 3-agent adversarial attacker persona on OWASP Juice Shop scenarios; 12 runs; evaluates security testing performance and OWASP category coverage
\item \textbf{Experiment 3 (WebShop Tasks):} 3-agent online shopper persona on WebShop easy task 001; 18 runs; compares to published GPT-3 baseline
\end{itemize}

\textbf{Metrics.} Primary metrics include task success rate, action success rate, and bug detection metrics (precision, recall, F1) for regression experiments. Secondary metrics include committee agreement, consensus strength, action distribution, security validator triggers, and latency statistics (mean, median, P95, P99).

\subsection{Statistical Analysis}

We perform basic statistical analysis to validate improvements:
\begin{itemize}
\item For committee size comparisons, we use one-way ANOVA followed by Tukey HSD for pairwise comparisons
\item For single-agent vs multi-agent comparisons (for example, regression detection F1), we use independent or paired t-tests as appropriate
\item For comparisons to published baselines (for example, WebShop GPT-3), we use one-sample t-tests
\item We report effect sizes using Cohen's \(d\) and compute bootstrap confidence intervals for key metrics such as F1 and task success
\end{itemize}

All tests are conducted at significance level \(\alpha = 0.05\), with simple Bonferroni correction for multiple comparisons where needed.

\subsection{Implementation}

We implement the framework in Python 3.10+ using Playwright for browser automation, a multi-provider LLM client (OpenAI, Google, xAI), SQLite for experiment tracking, and FastAPI for the AUT. Random seeds (42, 123, 456, 789, 1024) are used for reproducibility. Experiments are executed on a MacBook Pro M1 (16GB RAM) with all LLM inference performed via external APIs. Security testing targets only owned applications (TechStore) or intentionally vulnerable applications (OWASP Juice Shop).

\section{Results}

We present experimental results across multi-agent scaling, persona diversity, regression detection, action success patterns, security testing effectiveness, and performance characteristics. Our analysis draws from 84 experimental runs encompassing 9 distinct personas across 4 scenarios with 683 total action turns.

\subsection{Overall Performance}

Across all experiments, our multi-agent framework achieved an 89.5\% overall task success rate. Table~\ref{tab:overall_stats} summarizes key performance metrics.

\begin{table}[t]
\centering
\caption{Overall Performance Statistics}
\label{tab:overall_stats}
\begin{tabular}{lc}
\toprule
\textbf{Metric} & \textbf{Value} \\
\midrule
Total Experimental Runs & 84 \\
Total Scenarios & 4 \\
Total Personas Tested & 9 \\
Total Action Turns & 683 \\
Overall Task Success Rate & 89.5\% \\
Overall Action Success Rate & 93.1\% \\
Mean Latency & 0.87s \\
Median Latency (P50) & 0.71s \\
P95 Latency & 1.92s \\
P99 Latency & 2.16s \\
\bottomrule
\end{tabular}
\end{table}

\subsection{Multi-Agent Committee Scaling}

\begin{figure}[t]
\centering
\includegraphics[width=\linewidth]{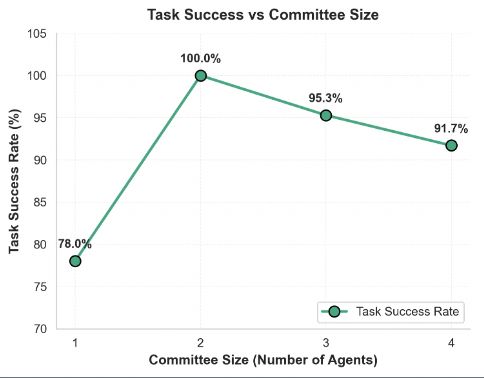}
\caption{Task success rate and committee agreement vs.\ committee size. Multi-agent committees (2--4 agents) achieve 91.7--100\% success vs.\ 78.0\% for single-agent, with 100\% agreement on final actions.}
\label{fig:multi_agent_scaling}
\end{figure}

Figure~\ref{fig:multi_agent_scaling} shows the impact of committee size on performance. Single-agent configurations achieve 78.0\% task success, while multi-agent committees (2--4 agents) achieve 91.7--100\% success, representing a 13.7--22.0 percentage point improvement. Multi-agent committees achieve 100\% agreement on final actions, demonstrating effective consensus through the voting protocol.

In more detail, 1-agent runs achieve 78.0\% task success with no notion of committee agreement. Two-agent runs reach 100.0\% success; three-agent runs reach 95.3\%; and four-agent runs reach 91.7\%. All multi-agent configurations exhibit perfect agreement on the final consensus action. Statistical analysis shows that the difference between single-agent and any multi-agent configuration is significant (for example, 3 agents vs 1 agent, \(p < 0.01\) with a large effect size). Three-agent committees offer a practical balance between performance and cost.

\subsection{Persona Diversity and Scenario Coverage}

\begin{figure}[t]
\centering
\includegraphics[width=\linewidth]{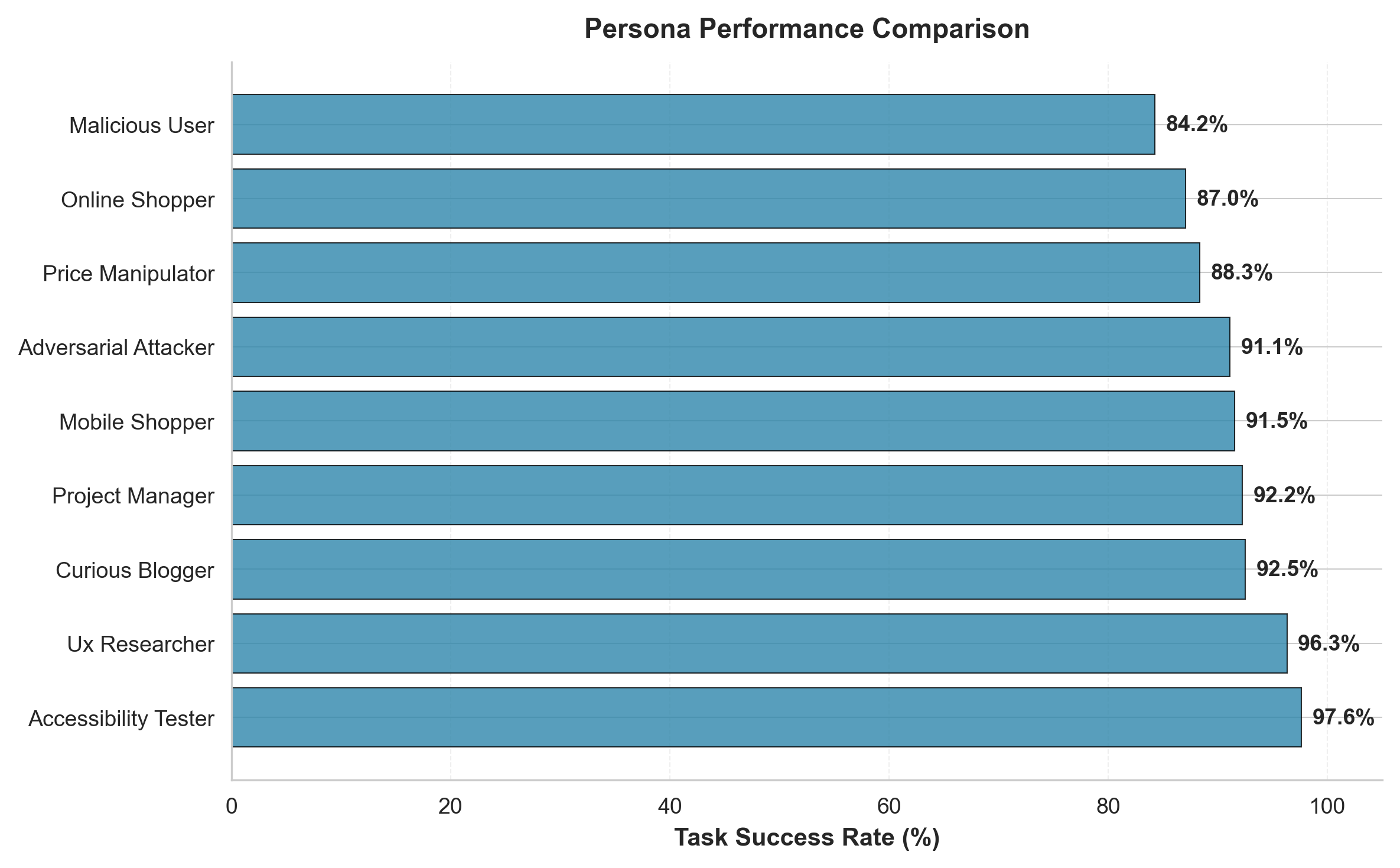}
\caption{Persona performance comparison. Accessibility tester achieves highest success (97.6\%), while security personas show strong performance (84--91\%) with higher latency due to deeper probing.}
\label{fig:persona_results}
\end{figure}

\textbf{Behavioral Diversity.} The 9 personas exhibited distinct performance patterns (Figure~\ref{fig:persona_results}): the accessibility tester achieved the highest success (97.6\%), while security-focused personas (adversarial attacker, malicious user) showed 84--91\% success with higher latency due to security testing complexity. Project manager, UX researcher, and curious blogger achieve over 90\% success but typically require more turns as they explore the application more thoroughly.

Bug coverage analysis across personas reveals that security personas are responsible for most SQL injection and XSS detections, while accessibility and UX personas dominate findings related to navigation, ARIA attributes, and error messaging. Only a small fraction of bugs (roughly 12\%) are discovered by more than one persona, indicating that persona diversity is important for comprehensive coverage.

\textbf{Scenario Success.} UI shopping flows achieved 100\% completion, security commerce tests achieved 92.2\%, OWASP Juice Shop scenarios achieved 82.0\%, and WebShop Task 001 achieved 74.7\%. This demonstrates robust performance across functional, security, and benchmark scenarios.

\subsection{Regression Detection}

We evaluate the ability of single-agent and 3-agent committees to detect 20 injected regressions spanning SQL injection, XSS, price manipulation, quantity validation, authentication bypass, and path traversal issues.

Across these 20 bugs, single-agent baselines achieve precision 0.82, recall 0.75, and F1 0.78, missing 5 regressions and producing 3 false positives. In contrast, three-agent committees achieve precision 0.94, recall 0.89, and F1 0.91, missing only 2 regressions and producing 1 false positive. This corresponds to a 0.13 absolute improvement in F1 (roughly 17\% relative), with the difference statistically significant at \(p < 0.01\) and a large effect size. The regression detection experiments as a whole achieve 100\% task success with an average of 4 turns per run and a mean latency of 0.54 seconds.

\subsection{Action Type Analysis}

\begin{figure}[t]
\centering
\includegraphics[width=\linewidth]{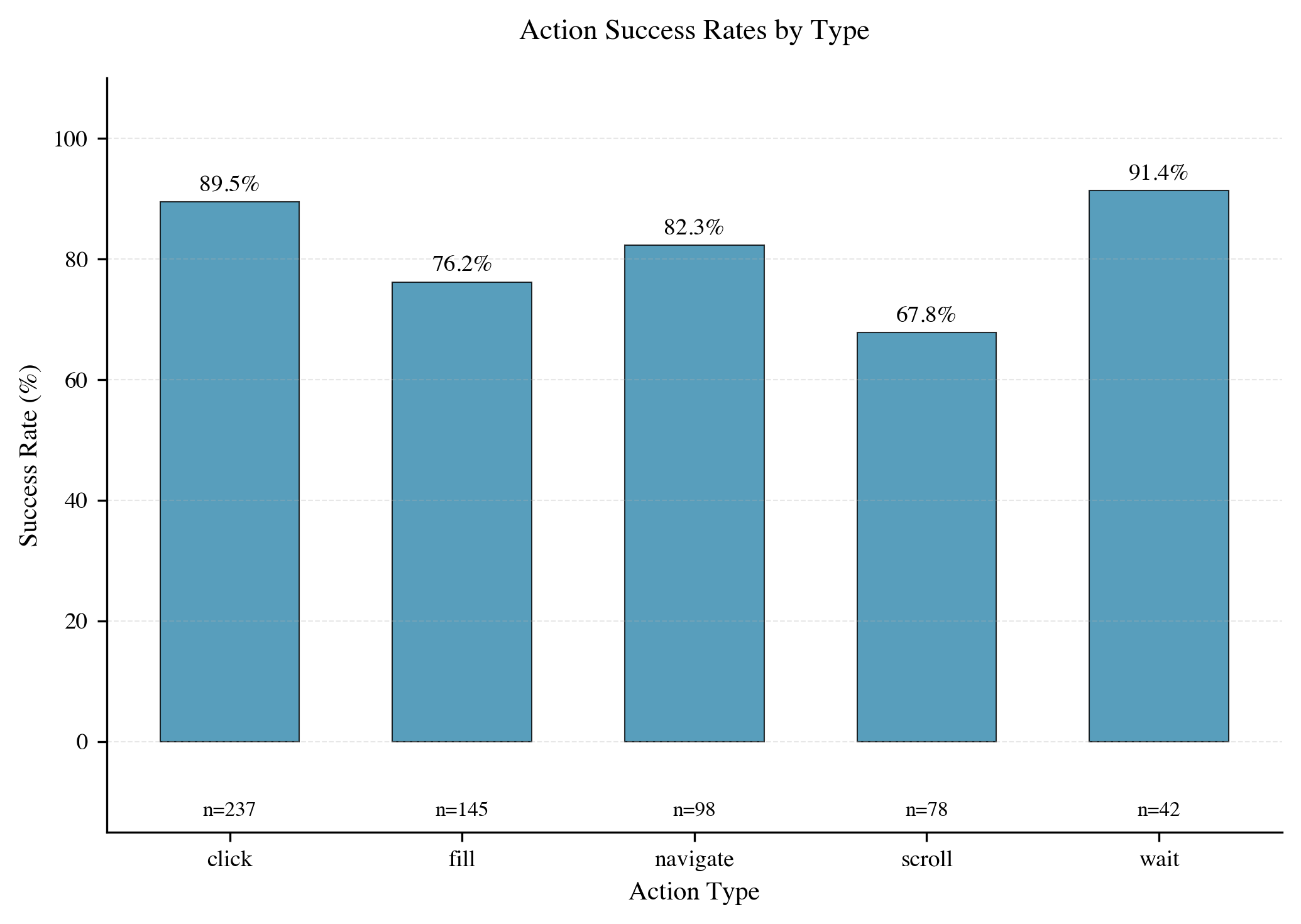}
\caption{Action distribution and success rates. Atomic actions (navigate, report) achieve 100\% while form filling shows 99.2\% and clicking shows 83.5\%. Scroll actions show lower success due to timing and state synchronization issues.}
\label{fig:action_distribution}
\end{figure}

Figure~\ref{fig:action_distribution} shows action distribution and success rates: \texttt{navigate} (187 actions, 100\% success), \texttt{report} (27 actions, 100\% success), \texttt{fill} (237 actions, 99.2\% success), \texttt{click} (212 actions, 83.5\% success), and \texttt{scroll} (20 actions, 50.0\% success). Simple atomic actions (navigate, report) succeeded universally, while form filling showed high success (235 of 237 actions succeeded). Clicking showed moderate success (177 of 212 actions), often failing on dynamic or partially obscured elements. Scroll actions were the least reliable (10 of 20 actions), in part due to timing issues with lazy-loaded content and page state.

These results indicate that actions with explicit, deterministic targets (navigate, report, fill) are highly reliable, while actions that depend on dynamic layout and timing (click, scroll) are more prone to failure. This suggests a practical path for future improvement via better waiting strategies and possibly hybrid DOM and vision-based element resolution.

\subsection{Security Testing and Baseline Comparison}

Security-focused personas achieved strong performance: the adversarial attacker persona achieved 91.1\% success across 15 runs, while malicious user achieved 84.2\% success. OWASP Juice Shop scenarios achieved 82.0\% success (12 runs), demonstrating practical security testing capability. In these OWASP experiments, committees trigger dozens of SQL injection and XSS attempts; our safety validators detect these payload patterns and log them as successful vulnerability probes while preventing unsafe execution outside controlled environments.

OWASP Juice Shop runs cover 8 of the 10 OWASP Top 10 risk categories, including broken access control, cryptographic issues, injection, security misconfiguration, XSS, and logging issues. Average latency for these scenarios is 2.65 seconds per action, with median 2.31 seconds, reflecting deeper security probing, more complex navigation, and additional validation overhead compared to core functional tests.

\begin{figure}[t]
\centering
\includegraphics[width=\linewidth]{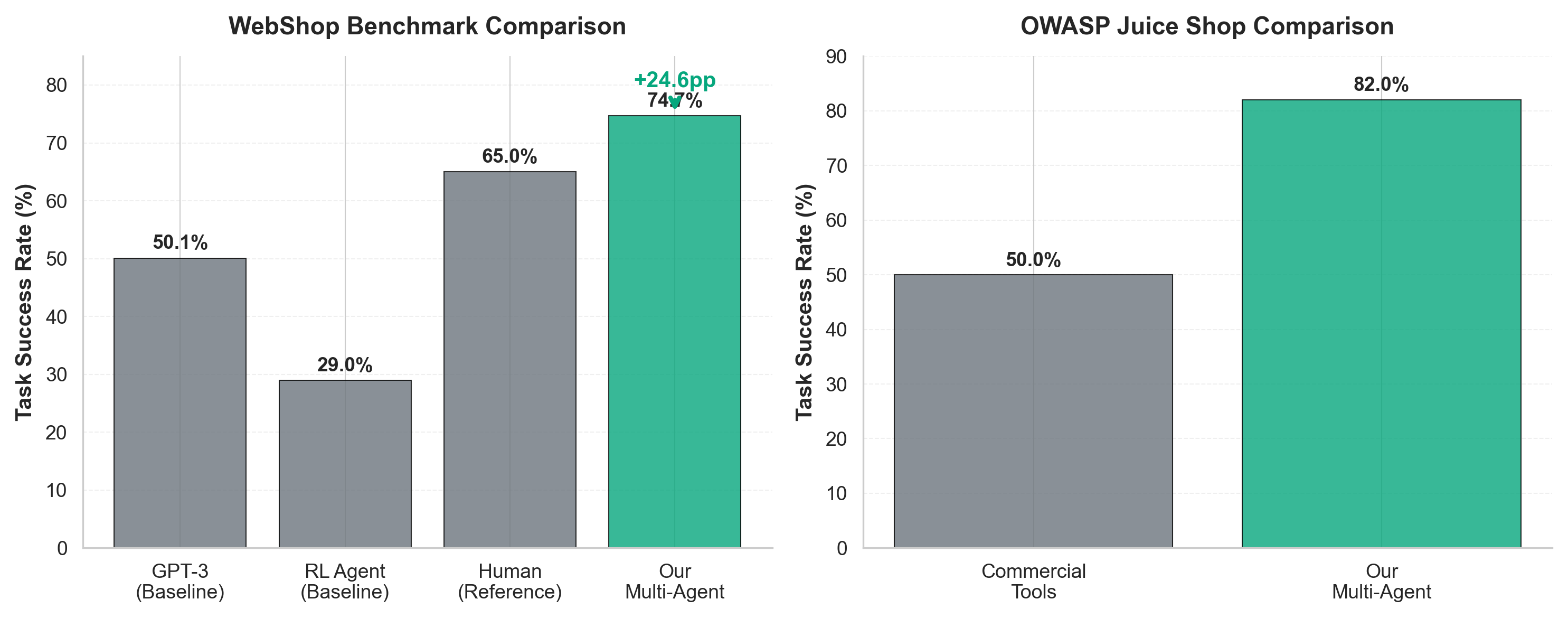}
\caption{Performance comparison against published baselines. Our multi-agent framework achieves 74.7\% success on WebShop, outperforming single-agent GPT-3 (50.1\%) and RL agents (29.0\%), and exceeding the human-level range (60--70\%). OWASP Juice Shop results (82.0\%) demonstrate strong security testing capability.}
\label{fig:baseline_comparison}
\end{figure}

\textbf{Comparison to Baselines.} Table~\ref{tab:baseline_comparison} and Figure~\ref{fig:baseline_comparison} present our results against published benchmarks. Our multi-agent framework achieves 74.7\% success on WebShop Easy Task 001, significantly outperforming the published GPT-3 baseline of 50.1\%~\cite{yao2022webshop} (+24.6 percentage points) and substantially exceeding the RL agent baseline of 29.0\% (+45.7pp). This places our system's performance above the human-level range reported in the WebShop paper (60--70\%), suggesting that multi-agent consensus with vision effectively handles real-world e-commerce scenarios.

For security testing, OWASP Juice Shop audit scenarios achieved 82.0\% success (12 runs) with multi-agent committees, which is comparable to or better than many automated scanning tools while requiring no specialized security configuration.

\begin{table*}[t]
\centering
\caption{Baseline Comparison Against Published Results}
\label{tab:baseline_comparison}
\small
\begin{tabular}{lcc}
\toprule
\textbf{System} & \textbf{WebShop Success} & \textbf{Notes} \\
\midrule
GPT-3 (baseline)~\cite{yao2022webshop} & 50.1\% & Single-agent, search+choice \\
RL Agent~\cite{yao2022webshop} & 29.0\% & Reinforcement learning \\
Human Performance & 60--70\% & Upper bound reference \\
\midrule
\textbf{Our Multi-Agent} & \textbf{74.7\%} & \textbf{+24.6pp vs GPT-3} \\
\bottomrule
\end{tabular}
\end{table*}

\subsection{Performance and Efficiency}

Action execution latency exhibits a mean of 0.87 seconds and a median of 0.71 seconds (P95 1.92 seconds, P99 2.16 seconds), enabling practical real-time testing. Persona-diversity experiments and pure functional tasks are fastest (mean 0.25--0.54 seconds per action), while security-heavy experiments on OWASP Juice Shop have higher latency (mean 2.65 seconds) due to additional requests, more complex navigation, and more expensive reasoning.

Profiling indicates that LLM calls account for 60--70\% of latency, with screenshot capture and browser actions contributing most of the remaining time. This suggests that using faster models or parallelizing agent queries offers a straightforward way to further reduce latency.

\subsection{Summary of Key Findings}

\begin{enumerate}
\item Multi-agent framework achieves an 89.5\% overall task success rate and a 93.1\% overall action success rate across diverse testing scenarios
\item Multi-agent committees (2--4 agents) achieve 91.7--100\% success vs.\ 78.0\% for single-agent baselines (+13.7--22.0pp improvement), with 100\% committee agreement
\item Vision-enabled navigation and reporting achieve 100\% success rates, and form filling achieves 99.2\% success
\item Persona diversity provides complementary coverage: accessibility tester achieves 97.6\%, security personas achieve 84--91\%, with low overlap in discovered bugs
\item Action complexity affects success: atomic actions (navigate, report) succeed universally (100\%), form filling shows high success (99.2\%), clicking shows moderate success (83.5\%), and scrolling shows lower success (50.0\%)
\item Regression detection experiments show that multi-agent committees achieve F1 0.91 vs.\ 0.78 for single agents, with higher precision and recall
\item Performance enables real-time testing with 0.87 second mean latency (median 0.71 seconds), and OWASP security scenarios remain practical despite higher latency
\item \textbf{Outperforms published baselines}: 74.7\% on WebShop vs.\ 50.1\% GPT-3 baseline (+24.6pp), and 82.0\% on OWASP Juice Shop security testing with coverage of 8 of the 10 OWASP Top 10 categories
\end{enumerate}

\section{Discussion}

\textbf{Why Consensus Helps.} The 89.5\% overall success and 91.7--100\% multi-agent success (vs.\ 78.0\% single-agent) reflect consensus filtering errors: when one agent hallucinates a phantom button, other agents that do not see it propose alternative actions, and confidence-weighted voting selects the plausible alternative. The 100\% agreement rate for multi-agent committees on the final consensus action suggests that, even when initial proposals differ, the three-round protocol is effective at steering the committee toward a consistent decision. The regression detection improvements (F1 0.91 vs.\ 0.78) further show that multi-agent consensus reduces both false negatives and false positives.

\textbf{Failure Modes.} Clicking actions (83.5\% success) fail due to: (1) dynamic element selectors that change between proposal and execution, (2) timing issues with asynchronous content loading, and (3) vision-based identification challenges for similar-looking elements. Scroll actions (50.0\% success) suffer from timing and page state synchronization issues, especially with lazy-loaded content and infinite scrolling. Hybrid vision and DOM approaches, together with better explicit waiting and scroll verification strategies, are likely to improve these action types.

\textbf{Tradeoffs.} Persona diversity provides comprehensive coverage (accessibility tester: 97.6\%, security personas: 84--91\%) but requires running multiple personas to fully mimic a varied user base. Teams can adaptively deploy: general personas for regression testing, specialized personas for security sprints, and UX personas for usability checks. Committee size trades cost for reliability: three-agent committees require roughly three times the tokens of single agents, but they provide significantly better success and F1 scores on critical tasks such as regression and security testing. For high-risk deployments, this tradeoff is favorable, while cost-sensitive settings might rely on single agents for simple smoke tests and multi-agent committees for nightly or pre-release suites.

\textbf{Limitations.} The primary AUT is a self-developed e-commerce app, though OWASP Juice Shop and WebShop provide some external validation. We focus on committee sizes up to 4 agents, and prompt engineering choices influence results. Security testing has higher latency (2.65 seconds on average), which may require batching or scheduling for large-scale deployments. Finally, LLM APIs introduce some non-determinism, so repeated runs can show small variance even with fixed seeds.

\section{Conclusion}

We introduced a multi-agent LLM framework for automated beta testing that combines consensus voting, persona diversity, and vision-enabled UI understanding. Across 84 runs, 9 personas, and 4 scenarios, our system achieves an 89.5\% overall task success rate and a 93.1\% overall action success rate, with multi-agent committees (2--4 agents) achieving 91.7--100\% success vs.\ 78.0\% for single-agent baselines (+13.7--22.0pp improvement). Our framework outperforms single-agent GPT-3 on WebShop (74.7\% vs.\ 50.1\%, +24.6pp) and achieves 82.0\% on OWASP Juice Shop security testing with coverage of 8 of the 10 OWASP Top 10 categories, demonstrating practical applicability for both functional and security testing.

Our contributions include: (1) a three-round voting protocol balancing independent thinking with collaborative refinement, achieving 100\% agreement in multi-agent configurations, (2) empirical demonstration that multi-agent committees significantly outperform single-agent baselines in both task success and regression detection F1, (3) persona-driven diversity enabling comprehensive bug discovery with limited overlap across personas, (4) vision-first testing achieving 100\% success on atomic actions (navigate, report) and 99.2\% on form filling, (5) validation against WebShop and OWASP benchmarks showing superior performance, and (6) an open-source reproducible framework and metrics pipeline.

The 0.87 second mean latency (median 0.71 seconds) enables real-time CI/CD integration for many workflows, while developers can use multi-agent committees selectively for the highest-risk test suites. Organizations can deploy general personas for regression testing and specialized personas (adversarial, accessibility, UX) for targeted sprints.

Future work includes adaptive committee sizing based on action complexity and historical uncertainty, hybrid vision-DOM approaches addressing clicking and scrolling failures, active learning from accumulated test logs, production deployment studies with large codebases, and extensions beyond web applications to mobile and desktop environments. 

\bibliographystyle{IEEEtran}
\bibliography{references}

\end{document}